\documentclass[11pt,a4paper]{article}
\usepackage{amsmath,amssymb}
\usepackage[pdftex,dvips]{graphicx}
\usepackage{color}

\newcommand{\jpsi}{\,\ensuremath{J/\psi}}
\newcommand{\ee}{\,\ensuremath{e^+e^-}}

\newcommand{\gevcc}{\,\ensuremath{\mathrm{GeV}/c^2}}
\newcommand{\mevcc}{\,\ensuremath{\mathrm{MeV}/c^2}}

\newcommand{\mev}{\,\ensuremath{\mathrm{MeV}}}
\newcommand{\cc}{\,\ensuremath{(c\bar{c})}}

\begin{document}

\title{\bf Recent results from Belle}
\author{T.~V.~Uglov$^a$\footnote{{\bf e-mail}: uglov@itep.ru}
\\
$^a$ \small{\em Institute for Theoretical and Experimental Physics} \\
\small{\em B. Cheremushkinskaya 25, Moscow, Russia}
}
\date{}
\maketitle

\begin{abstract}
The review on experimental results on charmonium and charmonium-like
spectroscopy from B-factories is presented.  Main theoretical
interpretations, such as conventional charmonium, molecular state,
hybrids, tetraquarks and others are discussed.
\end{abstract}

\section{Introduction}

Operation of two B-factories, Belle~\cite{belle} and
BaBar~\cite{babar}, is likely the most significant success in the
experimental particle physics of the last decade.  Their brilliant
work on CP-violation studies in B-mesons leads to Nobel Prize winning
by Kobayashi and Maskawa. In addition to the CPV studies B-factories
also made a huge contribution to the hadron spectroscopy, $\tau$-,
charm and two-photon physics.  One of the most prominent discovery
made by the B-factories is the observation of new charmonium-like
states, so called $X$, $Y$ and $Z$ mesons. In this paper an
experimental review of these states discovery and studies at
B-factories is presented, principal theoretical theoretical
interpretations are discussed.

\section{The Belle detector}

The Belle detector is a large-solid-angle magnetic spectrometer that
consists of a silicon vertex detector (SVD), a 50-layer central drift
chamber (CDC), an array of aerogel threshold Cherenkov counters (ACC),
a barrel-like arrangement of time-of-flight scintillation counters
(TOF), and an electromagnetic calorimeter (ECL) comprised of CsI(Tl)
crystals located inside a superconducting solenoid coil that provides
a $1.5\,$T magnetic field. An iron flux-return located outside the
coil is instrumented to detect $K_L^0$ mesons and to identify muons
(KLM). Two inner detector configurations were used. A beampipe at
radius $2.0\,$cm and a 3-layer silicon vertex detector were used for
the first sample of $\sim \! 156fb^{-1}$, while a $1.5\,$cm radius
beampipe, a 4-layer silicon detector and a small-cell inner drift
chamber were used to record the remaining data sample of $\sim \!
800fb^{-1}$.

\section{Charmonium: a long story}
\begin{figure}[t!]
\begin{center}
\includegraphics[width=\textwidth]{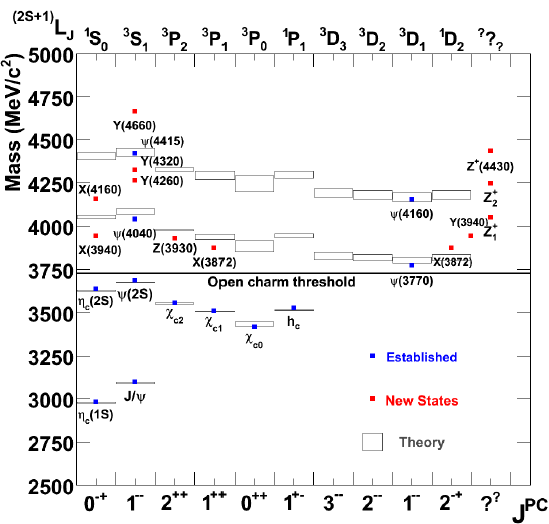}
\end{center}
\caption{Charmonium spectroscopy.}
\label {charmonium_spectroscopy}
\end{figure}

The charmed particles are known already more than for 35 years. The
first state, $\jpsi$ was observed by Ting and
Richter~\cite{TingRichter} in 1974, ten years after the $c$-quark
theoretical prediction. There was a great progress in charmonium
spectroscopy during next 6 years. Another 9 \cc\, states shown in
Fig~\ref{charmonium_spectroscopy} were discovered: $\psi(2S)$
$\eta_c$, $\chi_{c0}$, $ \chi_{c1}$, $ \chi_{c2}$, $\psi(3770)$,
$\psi(4040)$, $\psi(4160)$ and $\psi(4415)$.  Between 1980 and 2002
not a single new charmonium state was observed.  The beginning of the
new charmonium boom coincides with start of the B-factories operation.

There are several processes of  charmonium production at the B-factories:
\begin{enumerate}
\item production in B meson decays;
\item production in \ee\, annihilation through ISR photon;
\item double charmonium production in \ee\, annihilation;
\item production in two photon collisions.
\end{enumerate}
Some of these mechanisms allow to fix quantum numbers of the final
states. All charmonium states produced in \ee\, annihilation have
photon quantum numbers $J^{PC}=1^{--}$; \cc\, pairs from the
double charmonium production have opposite $C$-parities. Two photon
production fix $C=+1$ and forbids $J=1$.

Most of more than ten new charmonium-like states discovered at
B-factories do not fit the conventional charmonium spectroscopy scheme.
Observation of charmonium-like states with forbidden quantum numbers,
charged or extremely narrow width is the direct evidence of the
non-conventional spectroscopy. Thus the studies of the new states could
lead to the new revolution in the hadron physics.

\section{$\mathbf{X(3872)}$}

The first and may be the most mysterious charmonium-like particle,
$X(3872)$, was found by Belle collaboration in 2003~\cite{x3872} in
$B^+\to \jpsi\pi^+\pi^-K^+$ exclusive decay. Except for well-known
$\psi(2S)$ resonance, a new peak of more than $10\sigma$ significance
at mass $(3872.0\pm0.6\pm0.5)\mevcc$ was found in the
$\jpsi\pi^+\pi^-$ mass spectrum. Mass of the new state and presence of
\cc\, pair  in the decay products indicated
charmonium nature of $X(3872)$. However it's compatible with zero
width ($\Gamma_{tot}< 2.3\mev\,@\,90\,\%\,CL$) as well as mass equal
within errors to the sum of $D^{*0}$ and $D^0$ mesons rise doubts on
this interpretation. Until now it is unknown whether the mass of
$X(3872)$ is above or below $D^0D^{*0}$ threshold: $\Delta
m=-0.25\pm0.40\mevcc$. An existence and properties of $X(3872)$ was
confirmed by CD, D0 and BaBar collaborations~\cite{x3872conf}.

So narrow width for the state $~\sim 138\mev$ above the
$D\overline{D}$ threshold and absence of $X(3872)\to D\overline{D}$
decay ($\Gamma(X\to
D\overline{D})/\Gamma(X\to\jpsi\pi^+\pi^-)<7\,@\,90\%\,CL$~\cite{gammaDD_togammaJPSIPIPI})
means that the latter is forbidden either by parity conservation (if
$X(3872)$ has unnatural spin-parity $J^P=0^-,1^+,2^-,\dots$) or high
orbital momentum, {\it e.g.}  $J^P=3^-,4^+,\dots$ Presence of gluon or
pair of the light quarks except for \cc\,  pair in $X(3872)$
could also leads to suppression.

Observed enhancement at higher mass region in $m(\pi^+\pi^-)$
spectrum could be interpreted as $X\to\jpsi\rho^0$
decay~\cite{mpipi}. Since charmonium decay to $\jpsi\rho^0$ violates
isospin symmetry observation of this decay would be a strong argument
against charmonium nature of $X(3872)$ state. First evidence of
$X(3872)$ radiative decays was reported by Belle in
2005~\cite{radiativeXdecays_evidence} and confirmed by BaBar
collaboration in 2009~\cite{radiativeXdecays}. Decay
$X(3872)\to\jpsi\gamma$  was observed
with signal significance above $3.5\sigma$, thus
fixing $C(X)=+1$. Positive charge parity of $X$ supports hypothesis of
$\jpsi\rho^0$ decay.

Direct quantum numbers measurements are possible with angular
analysis.  Performed by the Belle collaboration~\cite{mpipi} it
excludes $J^{PC}=0^{++}$ and $0^{+-}$ possibilities. More accurate
analysis done by CDF~\cite{CDF_X_angular} excludes all $J^{PC}$ for
$J\le 3$, except for $1^{++}$ and $2^{-+}$.  Recent BaBar
studies~\cite{x_to_jpsi_omega} of $X(3872)\to\jpsi\omega$ decay
insignificantly favours negative parity assignment.

\section{$\mathbf{Y(3940)}$}

Another puzzling state, $Y(3940)$, was found by Belle
collaboration~\cite{y3940} in $B^+\to\jpsi\omega K^+$
decays. Near-threshold event excess in $\jpsi\omega$ spectrum was
interpreted as positive parity resonance of mass
$(3943\pm11\pm13)\mevcc$ and width $(87\pm22\pm26)\mev$.  Three years
later this state was confirmed by BaBar~\cite{babar_Y3940}, however
measured resonance mass and width, $(3914.6\pm2\pm1.9)\mevcc$ and
width $(34^{+12}_{-8}\pm6)\mev$, respectively, differs from previously
reported by Belle. Signal was found both for the charged and neutral B
meson decays with surprising ratio for this two channels
$N(B^\pm)/N(B^0)=0.27^{+0.28}_{-0.23}$, about $3\sigma$ below
predictions based on isospin symmetry. Observation of two-photon $Y$
production~\cite{Y_gammagamma} with mass and width similar to the
BaBar values fix $J_Y$ to $0$ or $2$. Quantum numbers of $Y$ could be
determined via angular analysis. These results will clear up the
nature of the resonance.

\section{Charmonium-like states in double charmonium production}

Double charmonium production is another source of charmonium-like
states. Studying recoil mass spectrum against \jpsi, Belle
collaboration found~\cite{X3940_belle}, among known charmonium states,
$\eta_c$, $\chi_{c0}$ and $\eta_{c}(2S)$, another narrow with respect
to the resolution $\Gamma_{tot}<52\mev\,@\,90\%\,CL$ peak of at mass
$(3943\pm6\pm6)\mevcc$. The discovery was soon confirmed by exclusive
process reconstruction in $\ee\to\jpsi D\overline{D}{}^*$
channel~\cite{X4160_belle}. Studies of similar final state $\jpsi
D^*\overline{D}{}^*$ leads to discovery of another narrow state of
width of $(139^{+111}_{-61}\pm21)\mev$ with mass
$(4156^{+25}_{-20})\mevcc$ decaying to $D^*\overline{D}{}^*$. Although
mass and width of new $X(4160)$ resonance agrees with those of known
$\psi(4160)$ charmonium state, charge parity, fixed by their
production mechanisms, is opposite.
 
\section{$\mathbf{J/\psi\varphi}$ resonances: $\mathbf{Y(4140)}$ \& $\mathbf{Y(4350)}$} 

Last year CDF reported evidence for the new narrow
state~\cite{CDF_Y4140} with mass of $4143.0\pm2.9\pm1.2\mevcc$ and
width $11.7^{+0.83}_{-5.0}\pm3.7\mev$. New resonance of $3.8\sigma$
significance was observed in $B\to \jpsi\varphi K$ decays of $B$
mesons produced in $p\bar{p}$ at $\sqrt{s}\sim
1.96\,\mathrm{TeV}$. Belle collaboration does not confirm this
peak~\cite{Belle_not_confirm}, however low efficiency near the
threshold disallows to report the contradiction. Search for this state
in $\gamma\gamma$ collisions also fails~\cite{gammagamma_belle_Y4350}.
At the same time a new narrow structure $\jpsi\varphi$ mass spectrum
of $3.2\sigma$ significance was observed at $4.35\gevcc$. Being
interpreted as resonance, this structure has mass and width of
$(4350.0^{+4.6}_{-4.1}\pm0.7\mevcc)$ and
$(13.3^{+17.9}_{-9.1}\pm4.1\mev)$, respectively. The significance of
both $Y(4140)$ and $Y(4350)$ states are less than $4\sigma$, thus
their confirmation is required. 

\section{ISR family: $\mathbf{Y(4008)}$, $\mathbf{Y(4260)}$, $\mathbf{Y(4350)}$ and  $\mathbf{Y(4660)}$}

First state of this family was discovered in 2005. Inspired by
$X(3940)$ observation, BaBar collaboration undertake a
search~\cite{babar_4260_discovery} for the resonance in
$\jpsi\pi^-\pi^+$ spectrum in the $\ee\to\jpsi\pi^-\pi^+\gamma_{ISR}$
process. A clear peak with mass and width of $(4264\pm12\mevcc)$ and
$(83\pm22)\mev$, respectively, was found. Similar analysis done by
Belle confirms $Y(4260)$, besides them another structure at mass of
$(4008\pm40^{+114}_{-28})\mevcc$ and width of $(226\pm44\pm87)\mev$
was found. Fit of the mass distribution with two interfering
Brait-Wigner shapes results in unavoidable splitting: there are two
solution of the same significance, resonance masses and widths, but
with different amplitudes.

In 2007 repeating the analysis for $\psi(2S)\pi^+\pi^-\gamma_{ISR}$
final state~\cite{babar_2007} BaBar found an evidence for a new
resonance decaying to $\psi(2S)\pi^+\pi^-$. Belle
confirms~\cite{belle_conf} this peak and found another one at higher
mass. The masses and widths of new $Y(4350)$ and $Y(4660)$ states were
determined to be
$M=4355^{+9}_{-10}\pm9\mevcc,\,\Gamma=103^{+17}_{-15}\pm11\mev$ and
$M=4661^{+9}_{-8}\pm 6\mevcc,\,\Gamma=47^{+17}_{-12}\mev$,
respectively. The latter states was also found in
$\Lambda_c\overline{\Lambda}{}_c$ mass spectrum in
$\ee\to\Lambda_c\overline{\Lambda}{}_c\gamma_{ISR}$ process~\cite{belle_lamlam}.

\section{Z: charged charmonium?}
Probably the most speculating charmonium-like states are those of
Z-family. The first evidence for charged charmonium-like state was
found by Belle in 2008~\cite{first_Z_plus}.  Studying $\psi(2S)\pi^+$
mass spectrum for $B\to\psi(2S)\pi^+K$ decay after discarding
$B\to\psi(2S)K^*$ events a narrow structure of $6.5\sigma$
significance near $4.43\gevcc$ was found.  Fit to this distribution
returns mass and width of the new $Z^+(4430)$ resonance of
$(4433\pm4\pm2)\mevcc$ and $(45^{+18}_{-13}{}^{+30}_{-13})\mev$,
respectively. Next year same data were analysed using more
sophisticated technique~\cite{first_Z_plus_repeat}. Fit to the Dalitz
distribution $M^2(\psi(2S)\pi^+)\, vs.\, M^2(K\pi^+)$ results in
similar $Z$'s mass and slightly higher width of
$(107^{+86}_{-43}{}^{+74}_{-56})\mev$. However, no significant signal
of $Z(4430)$ was found by the BaBar collaboration~\cite{Z_babar}: only
the upper limit have been set $\mathcal{B}(B\to Z^+K)\cdot
\mathcal{B}(Z^+\to\psi(2S)\pi^+)<3.1\cdot10^{-5}$, which is lower than
measured by Belle $(4.1\pm0.1\pm1.4)\cdot10^{-5}$.  Search for $Z$'s
in $B\to\chi_{c1}\pi^+K$ process turns out even more puzzling. Dalitz
plot fit to broad structure in $\chi_{c1}\pi^+$ mass results in two
charged charmonium-like resonances with masses of
$(4051\pm14^{+20}_{-41}\mevcc)$ and
$(4248^{+44}_{-29}{}^{+180}_{-25})\mevcc$, and widths of
$(82^{+21}_{-17}{}^{+47}_{-22})\mevcc$ and
$(177^{+54}_{-39}{}^{+316}_{-61})\mev$, respectively. A hypothesis of
two resonance is $\sim5.7\sigma$ more preferable in respect with one
$Z$.

\section {New states: interpretation}
There are several general interpretations of the new charmonium-like
states are now under discussion. Their current theoretical and
experimental status is summarized in Table~\ref{table}.

\begin{table}[h!]
\center
\begin{tabular}{|c|c|c|c|c|}
\hline
 State & $M$, \mevcc& $\Gamma$, \mev& $J^{PS}$ & Possible interpretation \\ 
\hline 
$X(3872)$ & $3871.4\pm0.6$ & $<2.3$  & $2^{-+}$  & Molecule, $\chi_{c1}^\prime$, $\eta_{c}(2S)$, tetraquark \\
$X(3915)^\dagger$ & $3914\pm4$ & $23\pm9$  & $0/2^{++}$  & $Y(3940)$ \\
$Z(3930)^\dagger$ & $3929\pm5$ & $29\pm10$  & $2^{++}$  & $\chi_c(2P)$ \\
$X(3940)$ & $3942\pm9$ & $37\pm17$  & $0^{+?}$  & $\eta_c(3S)$ \\
$Y(3940)$ & $3943\pm17$ & $87\pm34$  & $?^{?+}$  & Conventional \cc, hybrid \\
$Y(4008)$ & $4008^{+82}_{-49}$ & $226^{+97}_{-80}$  & $1^{--}$  & non-res $\jpsi\pi^+\pi^-$ \\
$X(4160)$ & $4156\pm29$ & $139^{+113}_{-65}$  & $0^{?+}$  &  $\eta_c(4S)$  \\
$Y(4260)$ & $4264\pm12$ & $83\pm22$  & $1^{--}$  &  $3^3D_1$  \\
$Y(4350)$ & $4361\pm13$ & $74\pm18$  & $1^{--}$  &  Molecule,tetraquark, hadrocharmonium, hybrid  \\
$X(4630)^\dagger$ & $4634^{+9}_{-11}$ & $92^{+41}_{-32}$  & $1^{--}$  & $Y(4660)$ \\
$Y(4660)$ & $4664\pm12$ & $48\pm15$  & $1^{--}$  &  Molecule,tetraquark, hadrocharmonium, hybrid  \\
$Z(4050)$ & $4051^{+24}_{-23}$ & $82^{+51}_{-29}$  & $?$  &  Molecule,tetraquark, hadrocharmonium, hybrid  \\
$Z(4250)$ & $4248^{+185}_{-45}$ & $177^{+320}_{-72}$  & $?$  &  Molecule,tetraquark, hadrocharmonium, hybrid  \\
$Z(4430)$ & $4433\pm5$ & $45^{+35}_{-18}$  & $?$  &  Molecule,tetraquark, hadrocharmonium, hybrid  \\
\hline
\end{tabular}
\caption {Charmonium-like states and their interpretation. Only states
  shown with $^\dagger$ sign are well interpreted.}
\label{table}
\end{table}

\begin{itemize}
\item {\it Conventional charmonium} is the most well known and the
  only experimentally proved model tuned to describe charmonium below
  the $D\overline{D}$ threshold. It constrains quantum numbers of the state: $J=L+S;
  P=(-1)^{L+1}; C=(-1)^{L+S}$.

  In conventional charmonium model $X(3872)$ state could be
  interpreted as $\chi_{c1}^\prime$. However the measured decay rate
  $\mathcal{B}(X\to\jpsi\gamma)/\mathcal{B}(X\to\jpsi\pi\pi)<0.2$ is
  much smaller than predicted ($\sim30$). Possible $X(3872)$ treatment
  as $\eta_c(2S)$ expects large $\Gamma(X\to gg)$ and tiny
  $\Gamma(X\to\jpsi\pi\pi)$ which contradicts to the experiment.

  Interpretation of $Y(3940)$ as conventional charmonium also faced
  the problems. Assuming that probability of $B\to Y(3940)K$ decay is
  less than $10^{-3}$, which is usual for $B\to \cc K;\,
  \cc=\eta_c, \jpsi, \chi_{c0},\chi_{c1},\psi(2S)$ decays, one could
  calculate $\Gamma(Y(3940)\to\jpsi\omega)\sim1\mev$, which is mote
  than 10 times large than for known charmonium transitions. Absence of
  $D\overline{D}$ and $D^*\overline{D}$ decays is also counts against
  charmonium interpretation.

  A lack of $D^{(*)}\overline{D}{}^{(*)}$ decays also prevents to the
  $Y(4260)$ interpretation as conventional $3^3D_1$ \cc\ state.

  For $X(3940)$ and $X(4160)$ , candidates for $3^1S_0$ and
  $4^1S_0$ states, respectively, the discrepancies between the
  predicted and measured masses reach $\sim100\mevcc$ and
  $\sim250\mevcc$, respectively.

  For two states observed in ISR, $Y(4360)$ and $Y(4660)$ there are no
  vacant $1^{--}$ states in the conventional charmonium spectroscopy
  scheme and they are too broad for it.

\item{\it Molecular state} is two mesons loosely bounded by pion or
  gluon exchange. 

  Molecule is the most popular interpretation of $X(3940)$. Belief
  that fine coincidence between $X$'s mass and $D^*\overline{D}$
  threshold is not accidental, observed decays to $D^*\overline{D}$ and
  a tiny radiative decay to \jpsi\ rate strongly supports molecular
  hypothesis. {\it Contra} arguments are too large $X\to
  \psi(2S)\gamma$ decay width and too small binding energy: $D$ mesons
  are too far from each other to be produced in $p\bar{p}$ collisions
  as an entity. However a possibility to mix with
  $\chi_{c1}^\prime$  state solves all these problems.
  
  No evidence for $Y(4140)$ production in two-photon collisions is
  found for now.  Small $\gamma\gamma$ width disfavours its molecular
  interpretation.

  Molecule composed of $D\overline{D}{}_1$ or $D_0\overline{D}{}^*$ is
  a good hypothesis for $X(3940)$ and $X(4160)$ states. 
  
  Charged $Z$'s could be treated as
  $D_{(s)}^{(*)}\overline{D}{}_{(s)}^{(*)}$ molecule.

\item{\it Tetraquark} is tightly bound four quark state.  
  
  Tetraquark interpretation predicts a list of new states with small
  mass spiting, especially for
  $(c\bar{u})(\bar{c}u),\ (c\bar{d})(\bar{c}d)$ and
  $(c\bar{u})(\bar{c}d)$ combinations. Intensive searches for
  $X(3872)$ find no evidence neither for charges nor for neutral
  partners.

  Some resonances from $Y$ family, $Y(4360)$ and $Y(4660)$ could be
  treated as $(\bar{c}q)(c\bar{q})$ tetraquark, $Y(4140)$ and
  $Y(4350)$ as $(s\bar{s}c\bar{c})$ diquark-antidiquark state. For
  $Z$'s there is $(c\bar{q}_1)(\bar{c}q_2)$ interpretation.

\item {\it Hybrid} --- meson with excited gluon degree of freedom.
  Interpretations of $Y(4360)$ and $Y(4660)$ states as hybrids is
  supported by lattice calculations, however in such a model, contrary
  to experiment, $D^{(*)}D^{**}$ decays should dominate.

\item {\it Hadrocharmonium}  --- charmonium coated by excited hadronic matter.
  $Y(4360)$ and  $Y(4660)$ could be treated as \cc and excited light meson hadrocharmonium.
  Excited charged light meson coating $\psi(2S)$ or $\chi_{c1}$ could give $Z$'s.

\item There are discussion on dynamical nature of new states.  {\it
  Threshold effects} caused by virtual states near the threshold could
  be responsible for these structures.

\end{itemize}

\section{Conclusions: a challenge}
Results of B-factories, Belle and BaBar, Tevatron experiments (CDF, D0)
as well as many others (BES etc.) start a new exciting era in the
hadron spectroscopy.  Seven years passed after discovery of $X(3872)$,
first charmonium-like state. More than ten another puzzling structures
were observed since that time and only a tiny fraction of them are
interpreted for now. New hadronic 'zoo' is a great challenge for
theoreticians to explain the nature of  these states and for
experimentalist to measure their properties with a highest possible
precision. New data from the upcoming Super-B-factories~\cite{superB}
would illuminate the mystery of this charmonium-like family and
hopefully solve XYZ puzzle.

\section{Acknowledgements}
This work is done with partial support of the Presidental grant
MK-4646.2009.2.



\begin{thebibliography}{00}
\bibitem{belle} A. Abashian {\it et al.}, Nucl. Inst. Meth. A {\bf 479}, 117 (2002).
\bibitem{babar}  B.~Aubert {\it et al.}, Nucl. Inst. Meth. A {\bf 479},  1 (2002).


\bibitem{TingRichter} J.~Aubert {\it et al.}, Phys. Rev. Lett.  {\bf 33}, 1402 (1974);
  J.~E.~Augustin {\it et al.}, Phys. Rev. Lett.{\bf 33}, 1406 (1974).
\bibitem{x3872} S.~K.~Choi {\it et al.}, Phys. Rev. Lett.  {\bf 91}, 262001 (2003).
\bibitem{x3872conf} D.~Acosta {\it et al.}, Phys. Rev. Lett.  {\bf 93}, 072001 (2004);
V.~M.~Abazov {\it et al.}, Phys. Rev. Lett.  {\bf 93}, 162002 (2004);
B.~Aubert {\it et al.}, Phys. Rev. D  {\bf 71}, 071103 (2005);
B.~Aubert {\it et al.}, Phys. Rev. D  {\bf 71}, 011101 (2005).
\bibitem{gammaDD_togammaJPSIPIPI} R.~Chistov {\it et al.}, Phys. Rev. Lett.  {\bf 93}, 051803 (2004).
\bibitem{mpipi}K.~Abe {\it et al.}, arXiv:hep-ex/0505038.
\bibitem{radiativeXdecays_evidence}K.~Abe {\it et al.}, arXiv:hep-ex/0505037.
\bibitem{radiativeXdecays} B.~Aubert {\it et al.}, Phys. Rev. Lett. {\bf 102}, 132001 (2009).
\bibitem{CDF_X_angular} A.~Abulencia {\it et al.}, Phys. Rev. Lett. {\bf 98}, 132002 (2007).
\bibitem{x_to_jpsi_omega}  B.~Aubert {\it et al.},  Phys. Rev. D  {\bf 82}, 011101 (2010).
\bibitem{y3940} S.~K.~Choi {\it et al.}, Phys. Rev. Lett.  {\bf 94}, 182002 (2004).
\bibitem{babar_Y3940} B.~Aubert {\it et al.}, Phys. Rev. Lett.  {\bf  101}, 082001 (2008).
\bibitem{Y_gammagamma} S~.Uehara {\it et al.}, Phys. Rev. Lett.  {\bf 104}, 092001 (2010).
\bibitem{X3940_belle} K.~Abe {\it et al.}, Phys. Rev. Lett.  {\bf 98}, 082001 (2007).
\bibitem{X4160_belle} P.~Pakhlov {\it et al.}, Phys. Rev. Lett.  {\bf 100}, 202001 (2008).

\bibitem{CDF_Y4140} T.~Aaltonen {\it et al.}, Phys. Rev. Lett.  {\bf 102}, 242002 (2009).
\bibitem{Belle_not_confirm} J.~Brodzicka Heavy Flavour
  Spectroscopy. XXIV International Symposium on Lepton Photon
  Interactions, Hamburg, Germany, August 17-22, 2009.
\bibitem{gammagamma_belle_Y4350} C.~P.~Shen  {\it et al.}, arXiv:0912.2383.
\bibitem{babar_2007}  B.~Aubert  {\it et al.}, Phys. Rev. Lett. {\bf 98}, 212001 (2007).
\bibitem{belle_conf} X.~L.~Wang {\it et al.}, Phys. Rev. Lett. {\bf99}, 142002 (2007).
\bibitem{belle_lamlam} G. Pakhlova  {\it et al.}, Phys. Rev. Lett. {\bf101}, 172001 (2008).

\bibitem{babar_4260_discovery}  B.~Aubert {\it et al.}, Phys. Rev. Lett. {\bf95}, 142001 (2005).


\bibitem{first_Z_plus}  S.~K.~Choi  {\it et al.},  Phys. Rev. Lett. {\bf100}, 142001 (2008).
\bibitem{first_Z_plus_repeat}  R.~Mizuk  {\it et al.},  Phys. Rev. D {\bf80}, 031104(R) (2009).
\bibitem{Z_babar}   B.~Aubert  {\it et al.},  Phys. Rev. D {\bf79}, 112001 (2009).



\bibitem{superB}  ``Letter of intent for KEK Super B Factory'', KEK Report 2004-4 (2004).



\end{thebibliography}
\end{document}